\documentclass[pre, notitlepage,longbibliography,superscriptaddress]{revtex4-1}

\usepackage{amsmath}
\usepackage{amssymb}
\usepackage{graphicx}
\usepackage{bm}

\usepackage[utf8]{inputenc}
\usepackage[T1]{fontenc}
\usepackage[english]{babel}
\usepackage{epstopdf}
\usepackage{natbib}

\usepackage[normalem]{ulem}
\usepackage[usenames,dvipsnames]{xcolor}
\usepackage{manfnt} 

\usepackage[colorlinks=false]{hyperref}

\newcommand \beq{\begin{equation}}
\newcommand \eeq{\end{equation}}
\newcommand{\upd}{\mathrm{d}}

\newcommand{\Across}{A_{\times}}
\renewcommand{\t}{h}
\newcommand{\FE}{I_E}

\renewcommand{\tt}{\tilde{t}}
\newcommand{\at}{\tilde{\alpha}}
\newcommand{\tmu}{\tilde{\mu}}
\newcommand{\ttperiod}{\tilde{T}}

\definecolor{MSLightBlue}{rgb}{.31,.506,.741}
\definecolor{MSDarkBlue}{rgb}{.51,.31,.741}
\definecolor{purple}{rgb}{0.5, 0.0, 0.5}

\begin{document}

\title{Dynamic buckling of an inextensible elastic ring: Linear and nonlinear analyses}

\author{Ousmane Kodio}
\affiliation{ Mathematical Institute, University of Oxford, Woodstock Rd, Oxford, OX2 6GG, UK}
\affiliation{Department of Mathematics, Massachusetts Institute of Technology, Cambridge, Massachusetts 02139, USA}
\author{Alain Goriely}

\author{Dominic Vella}\email[]{dominic.vella@maths.ox.ac.uk}
\affiliation{ Mathematical Institute, University of Oxford, Woodstock Rd, Oxford, OX2 6GG, UK}

\begin{abstract}
Slender elastic objects such as a column tend to buckle under loads.  While static buckling is well understood as a bifurcation problem,  the evolution of shapes during dynamic buckling is much harder to study. Elastic rings under normal pressure have emerged as a theoretical and experimental paradigm for the study of dynamic buckling with controlled loads. Experimentally, an elastic ring is placed within a soap film. When the film outside the ring is removed, surface tension pulls the ring inward, mimicking an external pressurization. Here we present a theoretical analysis of this process by performing  a post-bifurcation analysis of an elastic ring under pressure. This analysis allows us to understand how inertia, material properties, and loading  affect the observed shape. In particular, we combine direct numerical solutions with a post-bifurcation asymptotic analysis to show that inertia drives the system towards higher modes that cannot be selected in static buckling. Our theoretical results  explain experimental observations that cannot be captured by a standard linear stability analysis. 
\end{abstract}

\maketitle

\section{Introduction}
\label{sec:introduction}

Buckling is a ubiquitous instability in elastic materials \cite{tige61}. The \textit{static buckling} problem consists of determining the value of the load at which an instability takes place as well as the shape observed close to the instability. It  can be recast as a standard bifurcation problem for which there exists an extensive literature \cite{an05, Euler1744a,Levien2008}. The classic example of this is the Euler buckling of a column, which can be readily observed by attempting to compress a thin piece of paper along its axis \cite{Levien2008}: different buckling modes (eigenfunctions) exist, each with  a different buckling load (the eigenvalue).  In most situations it is the eigenfunction with the smallest eigenvalue that is observed experimentally; indeed, the higher modes are usually dynamically unstable, rapidly transitioning to the lowest mode \cite{Pandey2014}. 

Many other elastic structures  exhibit  a similar phenomenology:  For instance, the static buckling of an elastic ring under external pressure is a physical problem that has received much attention \cite{Levy1884,Carrier1947,Tadjbakhsh1967,Antman1965,Flaherty1973, Giomi2012,Biria2014,Chen2014, Fried2015}. These seminal works provide an in-depth analysis of the stability of the ring,  showing that the lowest mode is the `figure-of-eight' shape, determining higher modes and also offering different analytical and numerical methods through which to find the equilibrium shapes after bifurcation. These post-bifurcation equilibria are relevant to biological problems such as fluid flow through blood vessels \cite{Rubinow1972}. 

In contrast to the many studies on the static stability of elastic rings, the problem of \textit{dynamic buckling} has received much less attention. In dynamic buckling, the problem is to understand both the onset of instability under possibly dynamic loading and the evolution of the object's shape after the onset of instability. These problems are considerably harder as they involve both space and time \cite{Wah1970,cama84} and  simple energy arguments  cannot be used to obtain the post-buckling dynamics. For instance, twisted elastic rings going through a Michell instability \cite{Goriely2006} fold into a `figure-of-eight' shape without ever following equilibrium solutions. Dynamic loading can also be used to excite higher unstable modes that cannot be reached through static methods \cite{gota97,Audoly2005,Gladden05}. As such, dynamic buckling is a much richer phenomenon that offers many possibilities for new behaviors and, potentially, for the development of new active devices.

\begin{figure}[h!]
	\centering
	\includegraphics[width=0.6\linewidth]{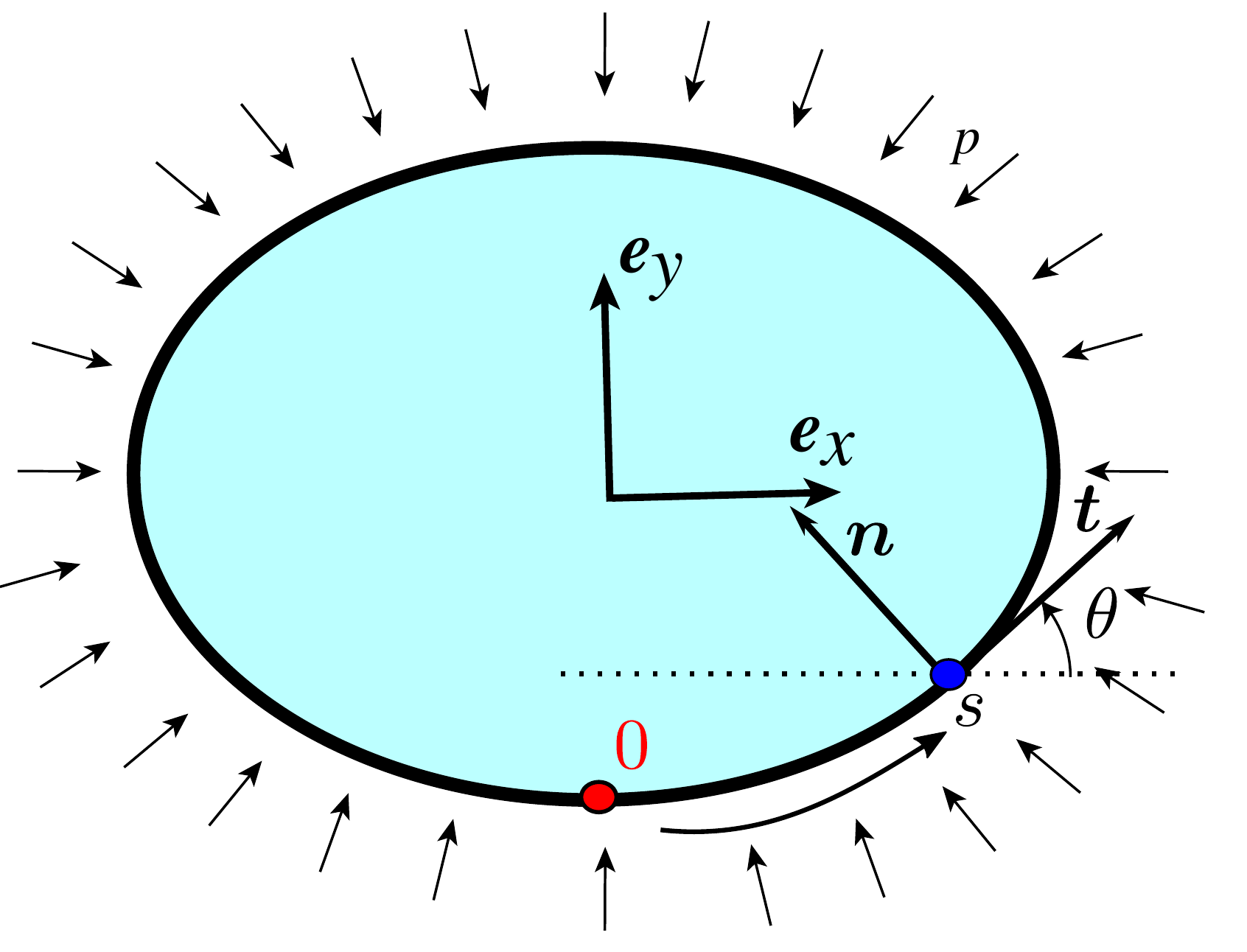}
	\caption{Schematic of an elastic ring subject to an external pressure $p$ (indicated by the arrows). The experiments described in a companion paper \cite{BKOCGV} achieve this via an unbalanced soap film filling the shaded area within the elastic ring (see Fig.~\ref{fig:ExptTimeSeries}).  The shape of the ring is described by its centreline with arc-length $s$ measured with respect to a point $0$. At each point $s$, the normal  and tangent vectors $\bm{n}$ and $\bm{t}$ are defined together with the angle $\theta$ between the horizontal $x$-axis and the tangent vector.}
	\label{fig:Diagram}
\end{figure}

In this paper we study the dynamic evolution of an elastic ring subject to a sudden jump in external pressure. When an elastic ring is subject to a slowly increasing external pressure, it buckles into its first unstable mode, a flattened, ellipse-like shape referred to as \textit{mode-2}. Higher modes with more lobes can be excited in dynamic loading and the problem here is to understand how these modes are selected and how they evolve after the onset of the instability.  In particular, we shall show that the presence of inertia implies that higher modes are selected, and we identify the mode number as a function of the driving pressure. Experimentally, these modes can be obtained by following the dynamics of an elastic ring embedded in a soap film, as presented in a companion paper \cite{BKOCGV}. Initially, the elastic ring is maintained in equilibrium by the balance of surface tension between an internal and external soap film. Once the external soap film is removed (by puncturing it at a point and allowing it to retract), the force from the internal soap film is suddenly unbalanced and pulls the ring inwards.  This is equivalent to an  external  pressure applied instantaneously to the ring, which is illustrated schematically in Fig.~\ref{fig:Diagram}. The experiments described in ref.~\cite{BKOCGV} demonstrate that a circular elastic ring buckling under surface tension in this way can exhibit a range of modes with higher modes selected dynamically as the importance of  surface tension  increases. An illustrative experimental time series of these experiments is shown in Fig.~\ref{fig:ExptTimeSeries}.

To develop some intuition for the problem of mode selection in dynamic buckling, we begin by considering the one-dimensional toy problem of an elastic beam (rather than a circular ring) of thickness, $\t$, under an in-plane compressive force ${\cal F}$ per unit width. In linear beam theory the governing equation for the vertical deflection $w(x,t)$ of a pinned-pinned beam of length $L$ is \cite{Lange1971}:
\begin{align}
&\rho \t \frac{\partial^2 w}{\partial t^2} + B \frac{\partial^4 w}{\partial x^4}+
{\cal F}\frac{\partial^2 w}{\partial x^2} = 0,\\
&w(-L/2)=w''(-L/2) = w(L/2)=w''(L/2)=0,
\label{eq:geq}
\end{align}
where $\rho$ is the density and $B$ is the bending stiffness of the beam. This equation exhibits growing buckled solutions of the form $w(x,t) =K \mathrm{exp}({\sigma t}) \cos[(2n+1)\pi x/L]$ if
\begin{align}
\rho \t \sigma^{2}  = \frac{(2n+1)^{2}\pi^2}{L^2}\left[ {\cal F} -B \frac{(2n+1)^{2}\pi^2}{L^2}\right].
\label{eq:dispersion}
\end{align}

By examining the maximum value of the growth rate, $\sigma$, as a function of $n$ in \eqref{eq:dispersion}, we see that the $n^{\mathrm{th}}$ mode is the fastest growing mode (and hence expected to be selected) if the compressive force is suddenly raised to ${\cal F}_{n}=  2(2n+1)^{2}\pi^2B/L^2$. Alternatively, if a particular force ${\cal F}$ is imposed then we might expect that the observed mode would be 
\beq
n\approx\frac{1}{2^{3/2}\pi}\left(\frac{{\cal F} L^2}{B}\right)^{1/2}-\frac{1}{2},
\label{eqn:nMaxSimp}
\eeq since it will be close to the fastest growing mode.  This  selection of a mode other than the lowest Euler mode (which corresponds to $n=0$) is made possible by the presence of the inertia term. 

Although the above calculation gives us an intuitive feel for the problem of dynamic buckling, it is expected to  be  only quantitatively accurate  for an elastic beam subject to a controlled compressive force ${\cal F}$. Achieving such a condition experimentally is difficult and is perhaps closest  to being achieved by an impactor with a known weight \cite{Gladden05}, though  then the imposed force becomes dependent on the imposed boundary conditions. The case of an elastic ring embedded in a soap film comes closer to a truly controlled force, but the question then becomes how the calculation that leads to \eqref{eqn:nMaxSimp} is modified by  the ring's curvature $\kappa=1/a$, with $a$ the ring radius. In a related study \cite{BKOCGV}, quantitative agreement with experiment was found by replacing the force ${\cal F}$ in \eqref{eqn:nMaxSimp} with the compressive force induced by the pressure difference $p$ combined with the ring curvature $1/a$ (Laplace's law), which gives ${\cal F}=pa$. In this paper we go beyond this simple physical argument to account for the effects of curvature completely, both in terms of the onset of instability and also in the post-buckling of the ring.

The paper begins in \S\ref{sec:theo} with a formulation of the governing equations from first principles,  together with our numerical technique and a presentation of the typical numerical results. To make analytical progress, \S\ref{sec:analysis} first re-formulates the governing equations of \S\ref{sec:theo} in a form that is more convenient for analytical work. This allows us to present a detailed linear stability analysis that goes beyond the simple-minded prediction of \eqref{eqn:nMaxSimp}. While the linear stability analysis is able to explain some features of our numerical results, other qualitative features of the numerics presented here and experiments of \cite{BKOCGV} are not explained by the solution of the linear problem. We therefore turn next to a weakly nonlinear analysis through which we derive an amplitude equation for the motion, which can be solved analytically. We illustrate the utility of this approach by applying the predictions of the weakly nonlinear analysis to explain various features of our numerical results, before summarizing our results in \S\ref{sec:conclusions}.

\section{Theoretical formulation\label{sec:theo}}

\subsection{Model problem}
We consider a circular elastic ring of radius $a$ subject to an externally applied pressure $p$.
We model the elastic ring as a rod of thickness $\t$ and length $L = 2\pi a$, whose shape is parametrized by its arc-length $s \in [0, L]$.  Previous work has considered the vibrations of arches, accounting for the effects of extensibility \cite{Nayfeh1995,Neukirch2012}. These works show that, under a controlled applied force, the effect of extensibility on the frequency of vibration is negligible provided that $\eta=h^2/(12L^2)\ll1$. The experiments of the companion to this paper \cite{BKOCGV} have $\eta\sim10^{-4}$ and are performed with controlled applied force. We shall therefore model the ring as being inextensible and unshearable to simplify the analysis somewhat. The ring is composed of a material of density $\rho$,  Young's modulus $E$, and its circular cross section has cross sectional area $\Across$ and second moment of area $I=\Across^2/(4\pi)$;  which gives a bending stiffness $B = EI=E \Across^{2}/(4\pi)$.

The position of a material point on the ring centreline at a time $t$ is denoted by the vector $\bm{r}(s,t)=x\bm{e}_{x}+y\bm{e}_{y}+z\bm{e}_{z}\in \mathbb{R}^{3}$ (Fig.~\ref{fig:Diagram}). Previous work has shown that the static equilibrium of a ring with an internal soap film may take a saddle shape \cite{Giomi2012,Chen2014}. However, in our problem the ring remains in the plane  and  we restrict our attention to planar deformations.  With the assumption of planarity, we denote by $\bm{t}$, $\bm{n}$ and $\bm{k}$ the  unit tangent, normal, and binormal vectors attached to the curve describing the centreline.  They satisfy the property $\partial_{s}\bm{r}(s,t)=\bm{t}$ and  $\partial_{s}\bm{t}=\kappa\bm{n}$, where $|\kappa|$ is the curvature. 

The  motion of the elastic ring is governed by Kirchhoff's equations \cite{LoveElasticity,AudolyPomeauBook}. Let $p \bm{n}$ be the external body force per unit length due to the applied pressure (integrated over the ring thickness) and denote by $\bm{F}$, $\bm{M}$  the resultant force and moment, respectively, acting on the centreline. The balances of linear and angular momenta then lead to \cite{an05,goriely17}:
\begin{align}
  &  \frac{\partial \bm{F} }{\partial s} + p\bm{n} = \rho \Across\frac{\partial^{2} \bm{r} }{\partial t^2 },\label{eq:Kirchhoff-Force} \\
  & \frac{\partial \bm{M}}{\partial s} + \bm{t}\times \bm{F}  = \bm{0}, \label{eq:Kirchhoff-Moment}
\end{align}
where, in eqn \eqref{eq:Kirchhoff-Moment}, we have ignored the rotary inertia terms \cite[p. 115]{goriely17}.

The governing equations are closed by the constitutive relation $\bm{M} = B\kappa(s,t) \bm{k}$ and the constraint that the ring is unshearable and inextensible.
We note that, alternatively, an energy formulation can  be used to establish the governing equations; for example,  the static buckling of twisted elastic rings within a soap film is studied in \cite{Biria2014,Fried2015}.

We denote by $\theta$ the angle between the $x$-axis and the tangent $\bm{t}$, so that $\bm{t}=( \cos \theta,\sin\theta)$. 
We rescale all lengths by the radius of the initial unstressed ring $a$, pressure by $p_{*}=B/a^3$, and time by the inertial time scale $t_{*} = a^2 \left( {\rho \Across}/{B} \right)^{1/2} $. In the new dimensionless variables, the governing equations projected on $\bm{e}_{x}$ and  $\bm{e}_{y}$,  are 
\begin{align}
& \frac{\partial x}{\partial s} = \cos \theta, \label{eq:elasticaSystemNonDim}\\
& \frac{\partial y}{\partial s} = \sin \theta, \label{eq:constraintY} \\  
& \frac{\partial^2 \theta}{\partial s^2} = F_{x} \sin\theta - F_{y} \cos\theta, \\
& \frac{\partial F_x}{\partial s} - P\sin\theta = \frac{\partial^2 x}{\partial t^2},\\
& \frac{\partial F_y}{\partial s} + P\cos\theta = \frac{\partial^2 y}{\partial t^2}. \label{eq:elasticaSystemNonDim_END}
\end{align}
This non-dimensionalization introduces the dimensionless parameter
\beq
P = \frac{p a^3}{B},
\label{eqn:Pdefn}
\eeq  which is the control parameter in the problem. It measures the importance of the work done by the external pressure compared to the bending energy of the ring, as described by Chen \& Fried \cite{Chen2014}.
We have the boundary conditions
\begin{align}
& \theta(2\pi,t) = \theta(0,t) +2\pi, \quad \partial_s\theta(2\pi,t) = \partial_s\theta(0,t),\quad
  x(2\pi,t) = x(0,t),	\\		&y(2\pi,t) = y(0,t), \quad  F_x(2\pi,t) = F_x(0,t),\quad 			 F_y(2\pi,t) = F_y(0,t),
\end{align}
and initially the ring is stationary. The initial shape of the ring is close to that of a circle of unit radius, though a small amount of noise is added to initiate instability in our numerical simulations.

\begin{figure}
	\centering
	\includegraphics[width=0.95\linewidth]{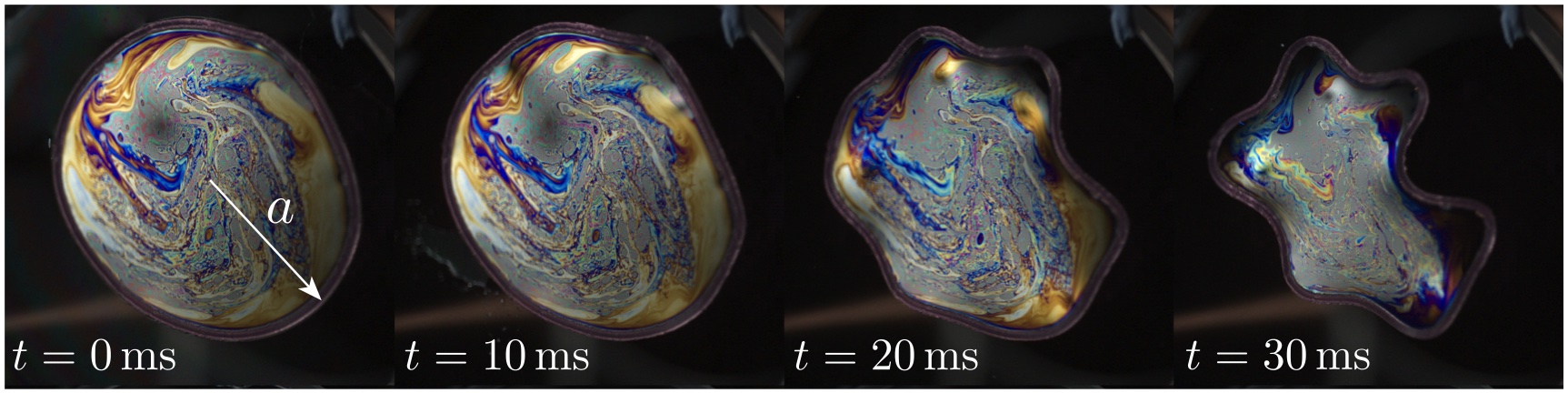}
	\caption{Time series from the companion paper \cite{BKOCGV} showing experimental images of the dynamic buckling of an elastic ring with square cross-section. This dynamic buckling is  achieved by placing an elastic ring (Young's modulus $E=42\mathrm{~kPa}$, thickness $\t=1\mathrm{~mm}$ and radius $a=24\mathrm{~mm}$) in a soap film and bursting the outer soap film at time $t=0$. This leaves an  unbalanced tension, $\gamma\approx26.5\mathrm{~mN/m}$, acting on the inner edge of the ring. This corresponds to an effective pressure difference $p=\gamma/\t\approx26.5\mathrm{~Pa}$ acting on the ring, or a dimensionless pressure, defined in \eqref{eqn:Pdefn}, $P\approx105$. Here, the characteristic time scale $t_\ast\approx0.31\mathrm{~s}$ so that the final image shown here corresponds to dimensionless time $t/t_\ast\approx0.1$. (Image courtesy of Finn Box.)}
	\label{fig:ExptTimeSeries}
\end{figure}

\subsection{Numerical analysis}

\begin{figure}
	\includegraphics[width=17cm]{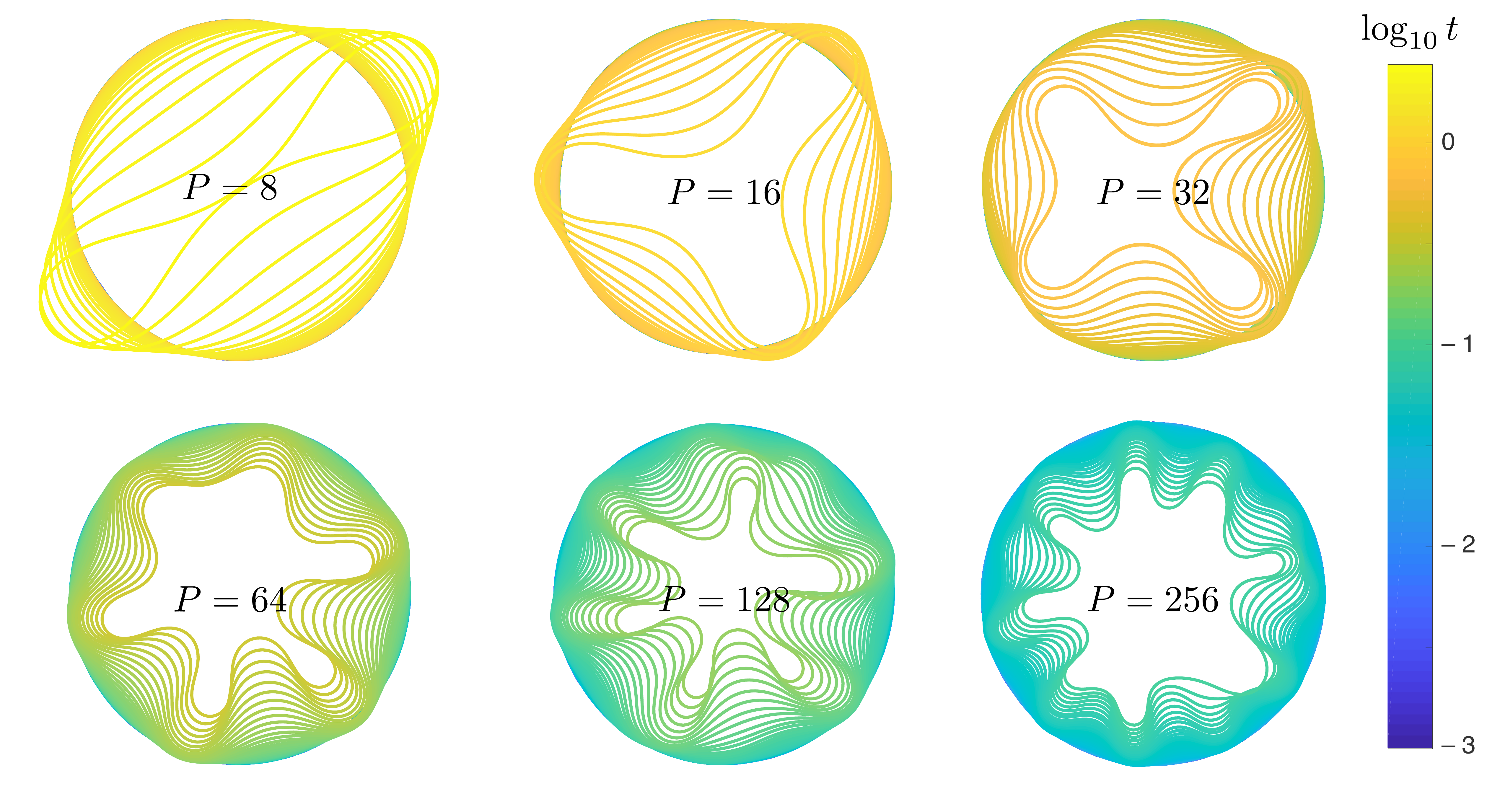}
	\caption{Dynamic evolution of an  elastic ring subject to a constant, externally-applied, pressure, determined from numerical simulations for $t>0$. Results are shown   for different values of the dimensionless pressure, $P$, defined in \eqref{eqn:Pdefn}. For each pressure, ring profiles are shown at different dimensionless times (with the time coded by its color as in the color bar on the right hand side).  Note that as the pressure increases (i) the observed mode number (the number of lobes) increases and (ii) the instability progresses more quickly (for the highest pressure, the instability has  progressed further, despite profiles being shown only for earlier times). These profiles are obtained by numerically solving the governing equations (\ref{eq:elasticaSystemNonDim}--\ref{eq:elasticaSystemNonDim_END}) starting with an approximately circular initial condition and zero initial velocity. These simulations use $N = 256$ grid points, a time step $\Delta t = 10^{-3}$, and a random perturbation to the initial circular state with $\varepsilon = 5\times 10^{-3}$. For a sense of the physical time and pressure scales, recall that the experimental snapshots shown in Fig.~\ref{fig:ExptTimeSeries} correspond to a physical pressure $p\approx26.5\mathrm{~Pa}$, dimensionless pressure $P\approx105$, and that the characteristic time scale there is $t_\ast\approx0.31\mathrm{~s}$.}
	\label{fig:NumericalZoology}
  \end{figure}

Before looking for asymptotic solutions, we present  numerical solutions for the evolution of this elastic ring under imposed pressure. 
We solve (\ref{eq:elasticaSystemNonDim}--\ref{eq:elasticaSystemNonDim_END}) numerically using a finite difference scheme in which  we discretize the  system  in space and time. This discretization produces a series of nonlinear equations, which are solved by using Newton's method at each time step (see Appendix A for details).  Note that we do not use a time-splitting projection method to enforce the inextensibility constraint; rather, our solution of the nonlinear equations  obtained by discretization, ensures that the geometrical relationships \eqref{eq:elasticaSystemNonDim} and \eqref{eq:constraintY} are satisfied, and hence that inextensibility is enforced at the same time. The initial velocity is zero, and the initial shape of the ring is circular with a uniformly distributed random number $\mathcal{R}_i\in[-\varepsilon,\varepsilon]$, with $\varepsilon\ll1$, added to the local radius of curvature at each point; in particular, the discretized initial state at arc-length $s_i=(i-1)\Delta s$ is
\begin{align}
\theta(s_i,0) = s_i , \quad x(s_i,0) = (1+\mathcal{R}_i)\sin(s_i),\quad y(s_i,0) = -(1+\mathcal{R}_i)\cos(s_i).
\label{eq:Init_Conditions}
\end{align}
Note that this initial condition is only consistent with the geometrical constraint~\eqref{eq:elasticaSystemNonDim}--\eqref{eq:constraintY} to  $O(\epsilon/(\Delta s))$. However, after the first time step, our numerical scheme ensures that the inextensibility constraint is  satisfied subsequently (see Appendix A for details).
 Snapshots of the evolution of the ring profile  for different values of the imposed pressure, $P$, are shown in Fig.~\ref{fig:NumericalZoology}. These show two trends: First, the selected mode number of the instability (defined as the number of lobes) increases with the imposed  pressure $P$. Second, the instability develops notably faster for larger values of  $P$. These initial observations will motivate our  asymptotic analyses of the problem.

\section{Analysis\label{sec:analysis}}

To facilitate the subsequent analysis, we first express the governing equations in the  moving frame of the ring $(\bm{t}, \bm{n}, \bm{k})$,  which is a right-handed orthonormal frame attached to the centreline of the ring at arc-length position $s$. We start by eliminating the force $\bm{F}$ from equation \eqref{eq:Kirchhoff-Force} to obtain a single equation for the vector position $\bm{r}$ together with the inextensibility constraint. The (dimensionless) bending moment is given by the constitutive equation $\bm{M} =  \kappa(s,t) \bm{k}$. Here, and henceforth, we use a prime to  denote derivatives with respect to the arc-length $s$ and  an overdot for  derivatives with respect to time $t$. Since $\kappa \bm{k}=\bm{r}' \times \bm{r}''$, the resultant moment can be written 
\begin{align}
\label{eq:Moment}
\bm{M} = \bm{r}' \times \bm{r}'' .
\end{align}
Substituting \eqref{eq:Moment} into the dimensionless version of \eqref{eq:Kirchhoff-Moment} then gives
\begin{align}
\label{eq:Fr}
\bm{r}' \times (  \bm{r}''' +\bm{F} ) = \bm{0},
\end{align} which immediately implies that
\begin{align}
 \bm{r}''' +\bm{F} = \lambda \bm{r}' + \beta \bm{k} ,
\end{align}
for some unknown functions $\lambda(s,t)$ and $\beta(s,t)$. However, since we restrict attention to planar deformations, we  have $\beta = 0$ and
\begin{align}
\label{eq:Force}
\bm{F} = - \bm{r}''' + \lambda \bm{r}'.
\end{align}
We note that, if redimensionalized, the term involving $\lambda$ in \eqref{eq:Force} would have the units of force and contributes to the tangential component of the internal force \cite{singh2018planar}.  Using the expression for the force \eqref{eq:Force} in \eqref{eq:Kirchhoff-Force} gives
\begin{align}
\ddot{\bm{r}} = - \bm{r}^{(4)} + (\lambda \bm{r}')' + P \bm{n},\quad \bm{r}'^2 = 1,\label{eq:GOE}
\end{align}
where the last equation is the local inextensibility constraint.
The two equations \eqref{eq:GOE} form a system for the two unknowns $\bm{r}(s,t)$ and $\lambda (s,t)$. 

Since we are interested in the evolution of the ring away from its initially circular configuration, we  take advantage of the  circular geometry to consider displacements from this shape with respect to the polar vectors associated with the circle. The curvature, $\kappa_{0}$, of the  ring in its initial configuration, together with the initial tangential  and normal vectors (denoted $\bm{t}_0$ and  $\bm{n}_{0}$, respectively), are:
 \begin{align}
& \kappa_{0} = 1, \\
& \bm{t}_{0} = \cos s\, \bm{e}_{x} + \sin s\, \bm{e}_{y} ,  \\
& \bm{n}_{0} = -\sin s\, \bm{e}_{x} +\cos s\, \bm{e}_{y}  .
\end{align}
Note that $\bm{t}' _{0} = \bm{n}_{0}$,  $\bm{n}' _{0} = - \bm{t}_{0}$ and the undeformed ring is parametrized by $\bm{r}_{0}  = - \bm{n}_{0}$.

We denote the displacements in the normal and tangential directions from the initial shape by $u(s,t)$ and $v(s,t)$, respectively. The ring shape may then be written as 
\begin{align}
\bm{r} = \bm{r}_0 + u \bm{n}_{0} + v \bm{t}_{0} .
\label{eq:R}
\end{align}
We  introduce  the angle $\phi(s,t)$ between the tangent of the deformed state and the tangent of the undeformed state  by  $ \bm{t} = (\cos\phi)\, \bm{t}_{0} + (\sin \phi)\, \bm{n}_{0} $. 
Taking the derivative of \eqref{eq:R} gives
\begin{align}
\label{eq:Rs}
\bm{r}'  =\bm{t}=(\cos\phi)\, \bm{t}_{0} + (\sin \phi)\, \bm{n}_{0}= \left[ 1 - u + v'  \right] \bm{t}_{0} + \left[ u' +  v \right] \bm{n}_{0}.
\end{align}
Hence, we  have:
\begin{align}
\label{eq:phi}
& \cos \phi = 1 -  u + v',  \quad \sin \phi = u' + v.
\end{align}
To relate the curvature to the angle $\phi$ we use $\bm{t}' = \kappa \bm{n}$, where $\bm{n} = -(\sin \phi) \bm{t}_{0} + (\cos\phi) \bm{n}_{0}$ so that
\begin{align}
\label{eq:Curvature}
\kappa = 1 + \phi' .
\end{align}
We see that the curvature has two components: a contribution, $\kappa_{0}=1$, coming from the fact that the ring is initially curved, and a contribution from the deformation of the ring relative to its initial circular shape.
Now taking the derivative of \eqref{eq:GOE} with respect to arc-length gives
\begin{align}
\ddot{\bm{t}} = -  \bm{t}^{(4)} + (\lambda \bm{t})'' - P \kappa \bm{t} . \label{eq:GOE-Tangent}
\end{align}

In the absence of pressure, namely $P = 0$, we recover the equation obtained by \citet{Burchard2003} for a closed loop elastica. (We note that in this particular case, local existence and uniqueness of the elastica solutions has been established for initial data in suitable Sobolev spaces, but that no global existence result exists \cite{Burchard2003}. Similarly, \citet{cama84} proved global existence for particular initial data in  a similar system that includes rotary inertial terms, but again with $P=0$.) We can resolve equation \eqref{eq:GOE-Tangent} into tangential and normal components giving a complete system of five equations for the five variables $(u,v,\lambda,\kappa,\phi)$ for the shape of an elastic ring subject to a normal pressure $P$.
\begin{align}
\label{eq:GOE-Polar-Start}
 &\cos \phi = 1 - u + \frac{\partial v}{\partial s},  \\
 &\sin \phi = \frac{\partial u}{\partial s} + v, \\
 &\kappa = 1 + \frac{\partial \phi}{\partial s}, \\
& \frac{\partial^{2} \phi}{\partial t^2} = -\left(  \frac{\partial^{3}\kappa}{\partial s^{3}}  -6 \kappa^2 \frac{\partial\kappa}{\partial s}  \right) + 2 \frac{\partial \lambda}{\partial s} \kappa + \lambda \frac{\partial\kappa}{\partial s} \\
&  \left(\frac{\partial \phi}{\partial t}\right)^2 =  \left[ \kappa^4 -3 \left(\frac{\partial\kappa}{\partial s}\right) ^{2} -4 \kappa \frac{\partial^{2}\kappa}{\partial s^{2}} \right]  - \frac{\partial^{2}\lambda}{\partial s^{2}} + \lambda \kappa^{2} + P \kappa. \label{eq:GOE-Polar-End}
\end{align}
We now make use of the formulation presented in eqns (\ref{eq:GOE-Polar-Start}--\ref{eq:GOE-Polar-End}) to perform a linear stability analysis of the problem; this is followed by a multiple scales analysis that allows us to examine the development of the instability beyond infinitesimal deformations (linear stability).

\subsection{Linear stability analysis}
\label{sec:subsection:LAS}

We begin by noting that  (\ref{eq:GOE-Polar-Start}--\ref{eq:GOE-Polar-End}) admit as a solution the undeformed circle: $v_{0} = u_{0} = \phi_{0} = 0$, $\kappa_{0} = 1$, provided that $\lambda_{0} = -P -1$. Physically, this corresponds to  the ring having an internal compressive force, $\lambda_{0} = -P -1$, which, combined with its initial curvature,  balances the external pressure (via Laplace's law). Our  interest lies in determining the  stability  of this equilibrium state.
Letting $u = u_{0} +\varepsilon u_{1}$, $v = v_{0} +\varepsilon v_{1}$, $\phi = \phi_0 +\varepsilon \phi_{1}$, $\kappa = \kappa_{0} +\varepsilon \kappa_{1}$, $\lambda = \lambda_{0} +\varepsilon \lambda_{1}$, with $\varepsilon\ll1$ arbitrary, we 
 expand (\ref{eq:GOE-Polar-Start}--\ref{eq:GOE-Polar-End})  to  first order in $\varepsilon$ to obtain:
 \begin{align}
   \label{eq:Linear-Order}
   & \frac{\partial v_{1} }{\partial s} = u_{1}, \\
   & \phi_{1} = v_{1} +\frac{\partial u_{1} }{\partial s},  \\
   & \kappa_{1} = \frac{\partial \phi_{1} }{\partial s},     \\
   & \frac{\partial^2  \phi_{1}}{\partial t^2} = (5-P)    \frac{\partial  \kappa_{1}}{\partial s}+ 2 \frac{\partial \lambda_{1} }{\partial s}  - \frac{\partial^{3} \kappa_{1} }{\partial s^{3}},   \\
   & 0=(P -2) \kappa_{1} + \lambda_{1} + 4 \frac{\partial^{2} \kappa_{1} }{\partial s^{2}}+ \frac{\partial^{2} \lambda_{1} }{\partial s^{2}}.
 \end{align}

 Seeking solutions of the form
 $u_1 = a_{1} \exp{(\sigma t + \text{i} n \theta)} $, $ v_1 = b_{1} \exp{(\sigma t + \text{i} n \theta)}$, $\kappa_1 =  c_{1}  \exp{(\sigma t + \text{i} n \theta)} $ ,
 $\phi_1 = d_{1}  \exp{(\sigma t + \text{i} n \theta)} $, $\lambda_1 = e_{1}  \exp{(\sigma t + \text{i} n \theta)} $,
 we obtain a homogeneous linear system that  has a non-trivial solution if, and only if, the associated determinant vanishes. This happens when
 \begin{align}
  \label{eq:Sigma}
  \sigma^{2} =  \frac{n^{2}(n^{2}-1)}{n^{2}+1}  \left[ P -( n^{2} -1) \right].
\end{align}
This dispersion relation may be recovered from that obtained previously by Wah \cite{Wah1970} if one neglects rotational inertia in their formulation (corresponding to the limit in which their parameter $\mu\propto a^2/\Across\to\infty$).  We also note that if $P = 0$ then $\sigma^{2}$ is negative so that $\sigma$ is
  imaginary and all the modes oscillate with a frequency given by the   imaginary part of $\sigma$, reproducing a result first obtained by Hoppe in 1871 \cite{Hoppe1871}.
  
\begin{figure}[!h]
	\centering
	\includegraphics[width=0.6\linewidth]{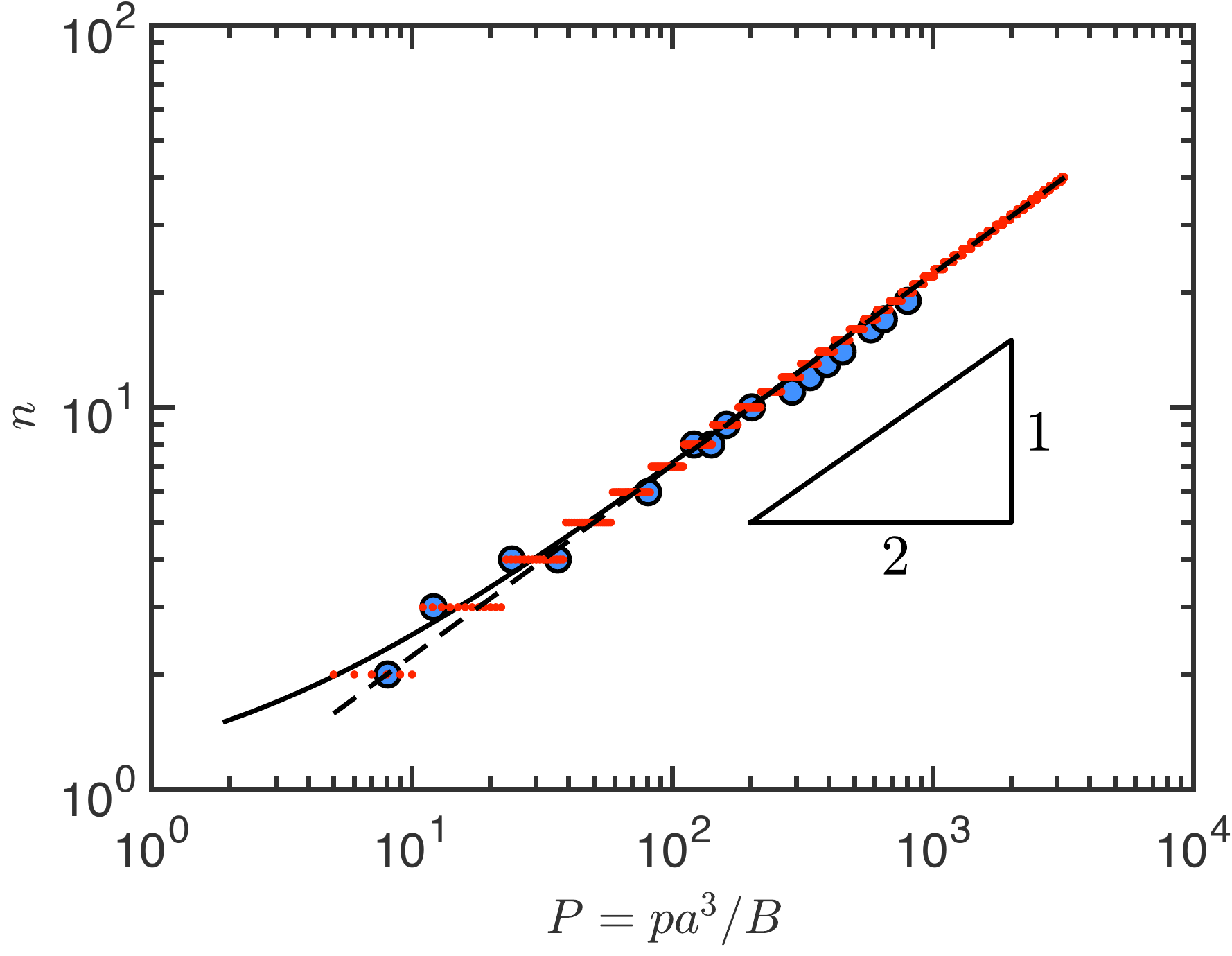}
	\caption{The mode number $n$ selected for different imposed external pressures, $P$:  large  blue dots correspond to the mode observed in numerical solutions of the full system (\ref{eq:elasticaSystemNonDim}--\ref{eq:elasticaSystemNonDim_END}). The mode number observed numerically agrees well with the most unstable (integer) mode predicted from a linear stability analysis (small red dots). This analytical result is itself well approximated by the  continuous prediction \eqref{eq:ModeSelection} (solid black curve) in which $n$ is treated as a continuous variable, while for $P \gg 1$ all approaches tend to the simple scaling  $n_{c} \sim (P/2)^{1/2}$, which is shown as the dashed line.}
	\label{fig:Results}
\end{figure}

  Our main interest lies in the instability that may be observed with positive external pressure, $P>0$. In this regard, note that  $\sigma = 0$ when
  $P =p_{n} = n^{2}-1$, giving another perspective on the $n^\mathrm{th}$ static buckling load of a ring under normal pressure  \cite{Tadjbakhsh1967, Flaherty1972}. For a given $P>0$, all  modes with  $n\leq (P+1)^{1/2}$ are unstable.  However, the mode that is expected to be most relevant in the development of instability from arbitrary initial conditions is the most unstable mode (the mode with the largest value of $\sigma$) for a given $P$. Since the mode number $n$ is an integer, the calculation of the most unstable mode number is obtained for a pressure interval as shown  in Fig.~\ref{fig:Results} (see  red dots). This value agrees well with the mode number observed in numerical simulations. Note that the numerically determined mode number is determined from the simulations at the point at which the instability is first visible in the shape. For large $n$ this appears to introduce an error in $n$ of 1, perhaps because one lobe disappears before instability has become numerically observable. An alternative, if approximate, approach is to treat $n$ as a continuous variable and obtain the value of $n$ that maximizes $\sigma$ in \eqref{eq:Sigma} using standard methods. This calculation reveals that the most unstable mode, $n_{c}$, is a solution of 
  \begin{align}
    \label{eq:ModeSelection}
    P = \frac{(n_c^{2}-1)(2 n_c^{4}+3 n_c^{2}-1)}{n_c^{4}+2 n_c^{2}-1},
  \end{align}
For $P \gg 1$, we find $n_c \sim (P/2)^{1/2}$, which in practice gives a very good account of the numerical results, and the more detailed analytical calculation, as shown by the dashed line in Fig.~\ref{fig:Results}. Note that this result is precisely that obtained by letting $L=2\pi a$ and ${\cal F}=pa$ in \eqref{eqn:nMaxSimp}: for large mode numbers (large dimensionless pressures) the most important role of the ring's curvature is to convert the externally applied pressure $P$ into an in-plane compression $\lambda$ within the ring (corresponding to ${\cal F}$ in the introductory beam analysis).

\subsection{Weakly nonlinear analysis}

The predictions of the linear stability analysis hold only for early times, $t\ll1$. As the amplitude of the perturbation to the circular state grows, various nonlinear terms compete with the exponentially growing terms found by solving the linear system; eventually, these nonlinearities can no longer be ignored. As a simple illustration of the importance of nonlinear effects, note that the linear theory predicts that the area enclosed by the ring is constant to leading order: $A(t)\approx\tfrac{1}{2}\int_0^{2\pi}(1+u)^2~\upd s=\pi+\mathcal{O}(\varepsilon^2)$. Moreover, if one naively calculates the correction at $\mathcal{O}(\varepsilon^2)$ using the linear expansion, one finds that the area enclosed should increase with time, in clear contradiction to the numerical results of Fig.~\ref{fig:NumericalZoology} and experimental results \cite{BKOCGV}. This prediction is a result of the inconsistency of using a result determined at $\mathcal{O}(\varepsilon)$ to make predictions about a correction at $\mathcal{O}(\varepsilon^2)$.

To go beyond the linear theory we perform a weakly nonlinear analysis. From  \eqref{eq:Sigma}, we see that as the external pressure increases, the ring remains stable as long as $\sigma^{2} \leq 0$ for
all $n \geq 0$. Therefore the ring is stable only  when $P \leq p_{n}$ for all $n \ge 2$ (with $p_{2} = 3$ the critical pressure above which the circular solution of the ring becomes unstable \cite{Tadjbakhsh1967}). We are interested in the dynamic evolution of a critical mode $n$ when the imposed pressure slightly exceeds  the critical pressure $p_{n}$. To find this evolution, we introduce a new small parameter, $\epsilon = \sqrt{P -p_n} \ll 1 $ (distinct from the arbitrary $\varepsilon$ used previously), to measure how far the pressure is above the critical pressure for a given mode. Therefore, we introduce $P=n^2-1+\epsilon^2$ in the equations that now depend explicitly on $\epsilon$. Next, we  expand all  variables to third order in $\epsilon$. For example, we write $u = u_{0} + \epsilon u_{1} + \epsilon^{2} u_{2} + \epsilon ^{3} u_{3} + \mathcal{O}(\epsilon^{4})$ with analogous expansions for the other variables $v,\kappa,\phi,\lambda$. These expansions are substituted into (\ref{eq:GOE-Polar-Start}-\ref{eq:GOE-Polar-End}) and the resulting hierarchy of linear systems  solved. Further details are given in Appendix \ref{app:wna}.

At  $\mathcal{O}(\epsilon)$, the solution is given by:
\begin{align}
\label{eq:U1}
&u_{1} = \alpha_1 \sin (n s),\quad v_{1} = -\frac{\alpha_1 }{n}\cos n s, \\
&\kappa_{1} = -\alpha_1 \left(n^2-1\right) \sin ns,\quad \phi_{1} = \frac{\alpha_1 \left(n^2-1\right) }{n}\cos n s,\\
&\lambda_{1}=\frac{\alpha_1 \left(n^2-1\right) \left(4 n^2-P+2\right) }{n^2+1}\sin ns,
\end{align}
where we have an arbitrary amplitude $\alpha_1$, that evolves on the long time scale $t=O(1/\epsilon)$.

At $\mathcal{O}(\epsilon^2)$, the solution is given by:
\begin{align}
\label{eq:U2}
&u_{2} = \frac{\alpha_1 ^2 \left(n^2-1\right)^2}{4 n^2},\quad v_{2} = -\frac{\alpha_1 ^2 \left(n^2-1\right)^2  }{8 n^3}\sin 2 n s ,\\
&\kappa_{2} =  - \frac{\alpha_1 ^2 \left(n^2-1\right)^2 }{4 n^2}\cos 2ns, \quad\phi_{2} = - \frac{\alpha_1 ^2 \left(n^2-1\right)^2 }{8 n^3}\sin 2ns, \\
&\lambda_{2} = \frac{3 \alpha_1 ^2 \left(n^4-1\right)^2 }{4 n^2 \left(n^2+1\right)}\cos 2ns.\label{eq:l2}
\end{align} 
Note that in the above expressions the  $\mathcal{O}(\epsilon^2)$ radial displacement,  $u_2$, is uniform, independent of arc-length $s$. This is contrary to the higher order radial displacement, $u_1$, which is oscillatory and, hence, does not contribute to a net radial displacement (once integrated over the whole ring). From this finding, we conclude that the clear visual impression that the ring shrinks radially (as seen in both the experiments \cite{BKOCGV} and numerical simulations of Figs \ref{fig:ExptTimeSeries} and \ref{fig:NumericalZoology}, respectively) is a second-order effect --- this striking feature of the problem can only be properly understood by going beyond the standard linear stability analysis.

The value of the amplitude $\alpha_1$ as a function of time is determined by a multiple-time expansion to derive the so-called amplitude equation. This equation is obtained as a compatibility condition for the existence of bounded solutions  by using the Fredholm alternative \cite{Lange1971,gonita99} for the linear system to third order (see Appendix~\ref{app:wna} for details). This condition reads
\begin{align}
\dfrac{\mathrm{d}^2 \alpha }{\mathrm{d}t^{2}} = \sigma^{2} \alpha - \mu \alpha^{3},
\label{eq:ampEq}
\end{align}
with
\beq
\alpha = \epsilon \alpha_1,\quad \sigma^{2} = \frac{n^2 (n^2 -1)}{n^2 +1} (P-p_{n}),\quad \mu =\frac{3}{8} \frac{ (n^2 -1)^4 }{n^2 +1}. 
\label{eqn:ampEqVars}
\eeq
If we neglect the non-linear term, $\mu \alpha^{3}$, we recover the results of  the linear stability theory, as expected, since \eqref{eqn:ampEqVars} gives the same linear growth rate $\sigma$ as \eqref{eq:Sigma}. 

To gain analytical insight into the behavior of the solution of the amplitude equation \eqref{eq:ampEq} we consider the particular problem of introducing a finite perturbation of size $\alpha(0)$ from rest, so that $\dot{\alpha}(0)=0$.  The solution of the linearized version of \eqref{eq:ampEq} would then be simply $\alpha(t)=\alpha(0)\cosh (\sigma t)$.  This linear solution, together with the form of \eqref{eq:ampEq}, suggests that we first introduce the rescaled variables $\tt=\sigma t$, $\at=\alpha/\alpha(0)$, which transform \eqref{eq:ampEq} to
\beq
\dfrac{\mathrm{d}^2 \at }{\mathrm{d}\tt^{2}} = \at - 2\tmu \at^{3},
\label{eq:ampEq2}
\eeq where
\beq
\tmu=\frac{\mu[\alpha(0)]^2}{2\sigma^2}
\label{eqn:tmuDefn}
\eeq is the sole remaining dimensionless parameter and measures the strength of the nonlinearity in the amplitude equation.

\begin{figure}
\includegraphics[width=0.6\linewidth]{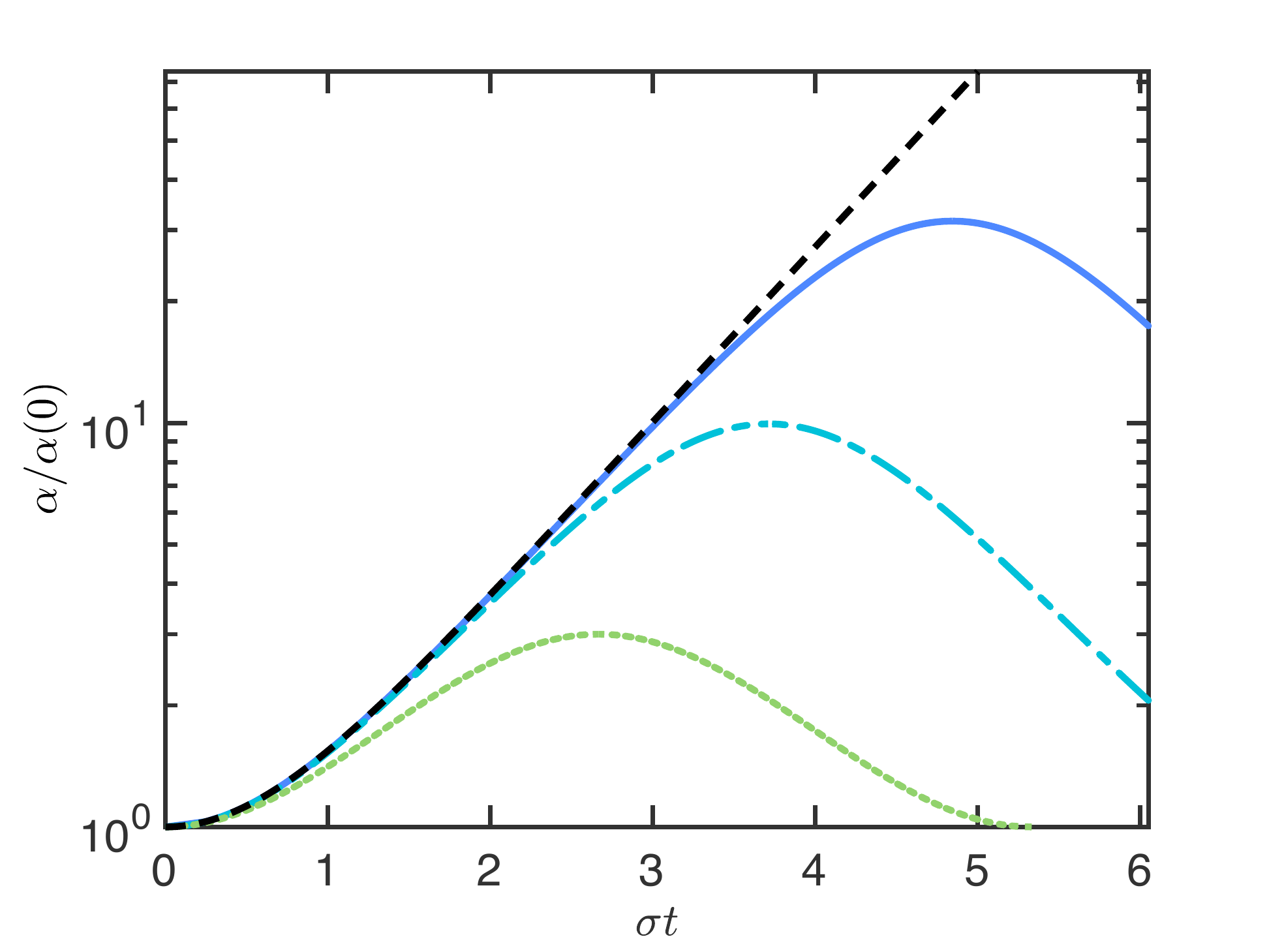}
\caption{The analytical solution of the amplitude equation \eqref{eq:ampEq2}, plotted from \eqref{eqn:EllipticIntegrals} with three different values of the dimensionless parameter $\tmu$, defined in \eqref{eqn:tmuDefn}. Results are shown for $\tmu=10^{-3}$ (solid curve), $\tmu=10^{-2}$ (dash-dotted curve) and $\tmu=10^{-1}$ (dotted curve) together with the prediction of the linear theory, $\alpha(t)/\alpha(0)=\cosh (\sigma t)$ (dashed curve). Larger values of the parameter $\tmu$ lead to earlier divergence of the amplitude equation from the linear prediction, i.e.~earlier onset of nonlinear effects, even when time is scaled by the linear growth rate $\sigma$.}
\label{fig:AmpEqnSoln}
\end{figure}

Eqn \eqref{eq:ampEq2} may readily be solved subject to the initial conditions $\at=1$ and $\upd\at/\upd\tt=0$ at $\tt=0$ to give 
\beq
\tt=(\tmu-1)^{-1/2}\left[\FE\left(\sin^{-1}\at,\frac{\tmu}{1-\tmu}\right)-\FE\left(\frac{\pi}{2},\frac{\tmu}{1-\tmu}\right)\right]
\label{eqn:EllipticIntegrals}
\eeq where $\FE(\phi,m)=\int_0^\phi(1-m\sin^2\theta)^{-1/2}\mathrm{~d}\theta$ is the Elliptic Integral of the first kind \cite{Olver2010}. The solution in \eqref{eqn:EllipticIntegrals} holds strictly only while $\at<[(1-\tmu)/\mu]^{1/2}$, which in turn requires $\tt<\ttperiod/2$ where the period of the motion
\beq
\ttperiod=2(\tmu-1)^{-1/2}\left\{\FE\left[\sin^{-1}\left(\sqrt{\frac{1-\tmu}{\tmu}}\right),\frac{\tmu}{1-\tmu}\right]-\FE\left(\frac{\pi}{2},\frac{\tmu}{1-\tmu}\right)\right\}.
\eeq
This analytical solution allows us to discuss briefly the effect of the nonlinearity.  We can see in Fig.~\ref{fig:AmpEqnSoln} the solution for three values of the parameter $\tmu$. As might be expected from the rescaled amplitude equation \eqref{eq:ampEq2}, the results are closer to the linear solution, $\at(\tt)=\cosh\tt$, when $\tmu$ is small. More quantitatively, the nonlinear term may be neglected while $\at\ll\tmu^{-1/2}$, i.e.~while $\tt\ll\cosh^{-1}(\tmu^{-1/2})\sim -(\log \tmu)/2$. In particular, since for large $n$, $\sigma\sim n(P-p_n)^{1/2}$ and $\mu\sim 3n^6/8$ so that $\tmu\sim 3n^4[\alpha(0)]^2/[16(P-p_n)]$, we expect that the role of nonlinearities will become more important more quickly for larger imposed pressures. (Note that this sense of `faster' onset of nonlinearity is measured in the natural time variable $\tt=\sigma t$, and so goes beyond the prediction of linear stability analysis that the linear growth rate of instability also increases with pressure.)

\subsection{Applications of the weakly nonlinear analysis}

Since  the amplitude $\alpha$ may readily be computed as a function of time (either using the above expressions or by direct numerical solution of the amplitude equation), we are now in a position to compare directly the numerical solutions of the full system with the post-bifurcation solution to second-order that is provided by our weakly nonlinear analysis. We study several aspects of the problem that are accessible in the full numerical solutions to illustrate the application of our weakly nonlinear analysis.  Before doing so, we note  that the shape of the ring is related to the radial and tangential displacements ($u$ and $v$, respectively) via \eqref{eq:R}, which may be written in component form as
\begin{align}
	& x(s,t) = \bigl[1-u(s,t)\bigr]\sin s + v(s,t) \cos s, 	\label{eqn:WNA_ShapeX}\\
	& y(s,t) = \bigl[u(s,t)-1\bigr]\cos s + v(s,t) \sin s.
	\label{eqn:WNA_ShapeY}
\end{align} As a result, the displacement field can easily be inferred from the shape, and \emph{vice versa}.

To facilitate the comparison between the prediction of the weakly nonlinear analysis and our numerical solution of the problem, we ensure that we use the same initial conditions in both problems; in particular, we solve \eqref{eq:elasticaSystemNonDim}--\eqref{eq:elasticaSystemNonDim_END} numerically  with an initial condition seeded by the weakly nonlinear analysis with a particular mode number, \textit{i.e.}~we use \eqref{eqn:WNA_ShapeX}--\eqref{eqn:WNA_ShapeY} with $u$, $v$ being the perturbation solution provided by \eqref{eq:U1}--\eqref{eq:l2} at $t=0$,
\begin{align}
&x(s,0) = \sin s-\alpha(0) \left(\sin s \sin n s+\frac{\cos s \cos ns}{n}\right)-
   \alpha(0)^2 \left[\frac{\left(n^2-1\right)^2 }{4n^2}\sin s+\frac{\left(n^2-1\right)^2 }{8 n^3}\cos s \sin 2ns\right]\\
&y(s,0) = -\cos s+ \alpha(0) \left(\cos s \sin ns-\frac{\sin s \cos ns}{n}\right) + \alpha(0)^2 \left[\frac{\left(n^2-1\right)^2 }{4
   n^2}\cos s-\frac{\left(n^2-1\right)^2 }{8 n^3}\sin s \sin2ns\right]
   \label{eqn:WNA_IC}
\end{align} 
rather than the random initial condition \eqref{eq:Init_Conditions} used for computations with fixed pressure, but $n$ not known \emph{a priori}. Here, we choose $\alpha(0) = 0.01$.

\begin{figure}
\includegraphics[width=\linewidth]{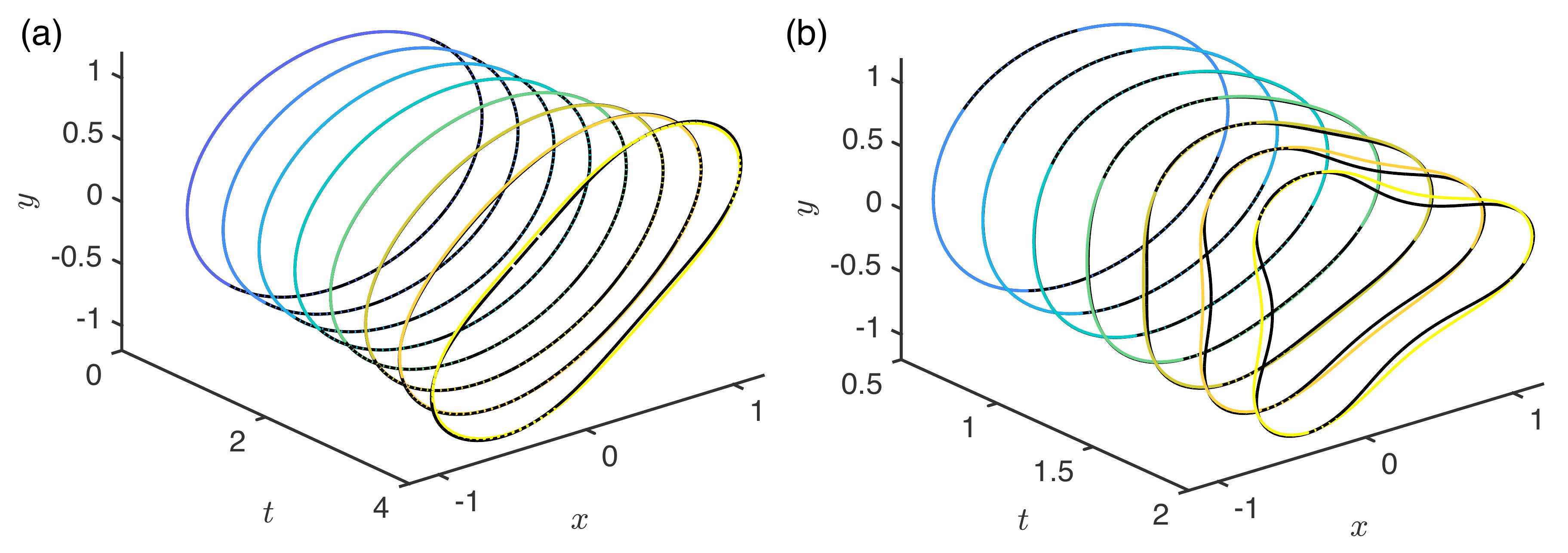}
\caption{Comparison between the numerical solution (solid curves) and the weakly non-linear solution (dotted curves) for (a) $P=3.5$, corresponding to mode 2, and (b) $P = 8.7$, corresponding to mode 3. Shapes are shown at instants of time that start with $t=0.5$ and increase at constant intervals of time $\delta t$; in (a) $\delta t=0.5$ while in (b) $\delta t=0.25$. }
\label{fig:MSA-Shape}
\end{figure}

\paragraph{Shape} We compare the shape obtained by numerically solving the governing equations with that predicted by the weakly nonlinear analysis. The predicted shape is reconstructed from the amplitude $\alpha(t)$ and using the predictions \eqref{eq:U1} and \eqref{eq:U2}.  In detail, the profile predicted by  the weakly nonlinear analysis may be constructed as follows: using \eqref{eq:U1} and \eqref{eq:U2},  the displacement field $(u,v)$ is given by $u=u_0+\epsilon u_1 + \epsilon^2 u_2$ and $v=v_0+\epsilon v_1 + \epsilon^2 v_2$, which may then be directly substituted into \eqref{eqn:WNA_ShapeX}--\eqref{eqn:WNA_ShapeY}. The results of this reconstruction are shown in Fig.~\ref{fig:MSA-Shape} together with the numerical results for $P=3.5$ and $P = 8.7$, which correspond to the  $n=2$ and $n=3$ modes, respectively. The results of this comparison are very favourable.
\\

\begin{figure}[!h]
	\centering
	\includegraphics[width= 0.95\linewidth]{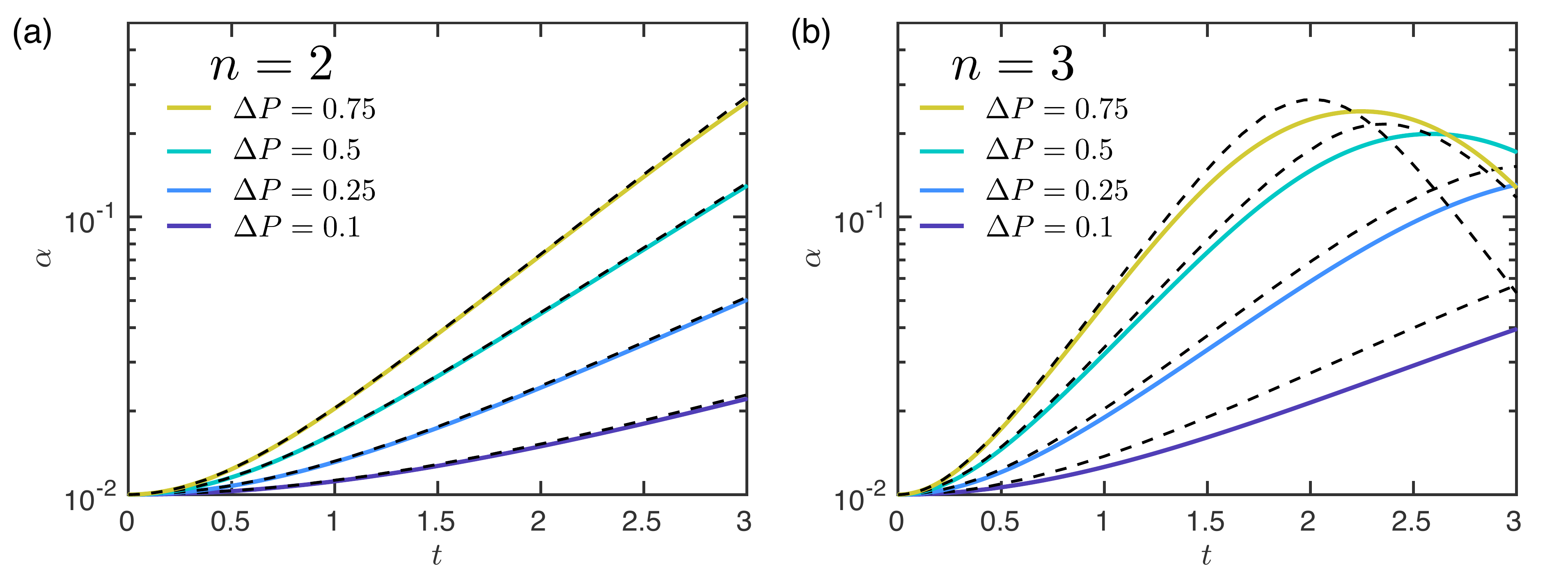}
	\caption{ Evolution of the amplitude $\alpha(t)$ determined as a function of time from full numerical simulations, using \eqref{eqn:AlphaNum} (solid curves) together the corresponding prediction of the amplitude equation \eqref{eq:ampEq} (dashed curves). Results are shown for (a) $n = 2$ and (b) $n=3$ with pressures  $\Delta P=P-p_c(n)=0.1, 0.25,0.5 $ and $0.75$ coded by color, as in the legend. (Recall that the critical pressure to excite mode $n$ is $p_c(n)=n^2-1$.)}
		\label{fig:A1}
\end{figure}

\paragraph{Amplitude} We  compare the amplitude $\alpha(t)$  with that obtained from the numerical simulations. To do so, we compute numerically the function $u(s,t)$ and extract the amplitude of its first Fourier component:
\beq
\alpha_{\text{num}}(t)= \frac{1}{\pi} \left[\left(\int_{0}^{2\pi} u(s,t) \sin n s\, \text{d}s\right)^{2}+\left(\int_{0}^{2\pi} u(s,t) \cos n s \,\text{d}s\right)^{2}\right]^{1/2}.
\label{eqn:AlphaNum}
\eeq
We see from Fig.~\ref{fig:A1} that the amplitude equation  captures quantitatively the evolution of the $n=2$ mode but only qualitatively captures the evolution of the $n=3$  mode.

\paragraph{Area} From the post-bifurcation solution to second order, we can compute the area of the ring including a perturbation with mode number $n$. We find that this area is
\begin{align}
A_n(t) = \pi - \frac{(n^2-1)\pi}{2} \alpha(t)^2.
\end{align} The change in area, $\Delta A(t)=A(0)-A_n(t)=\pi-A_n(t)$ and the relative change in area is 
\begin{equation}
\frac{\Delta A(t)}{A(0)}=1-\frac{A_n}{\pi}=\frac{n^2-1}{2}\alpha(t)^2.
\label{eqn:DeltaAPredMSA}
\end{equation}

\begin{figure}[!h]
	\centering
	\includegraphics[width= 0.95\linewidth]{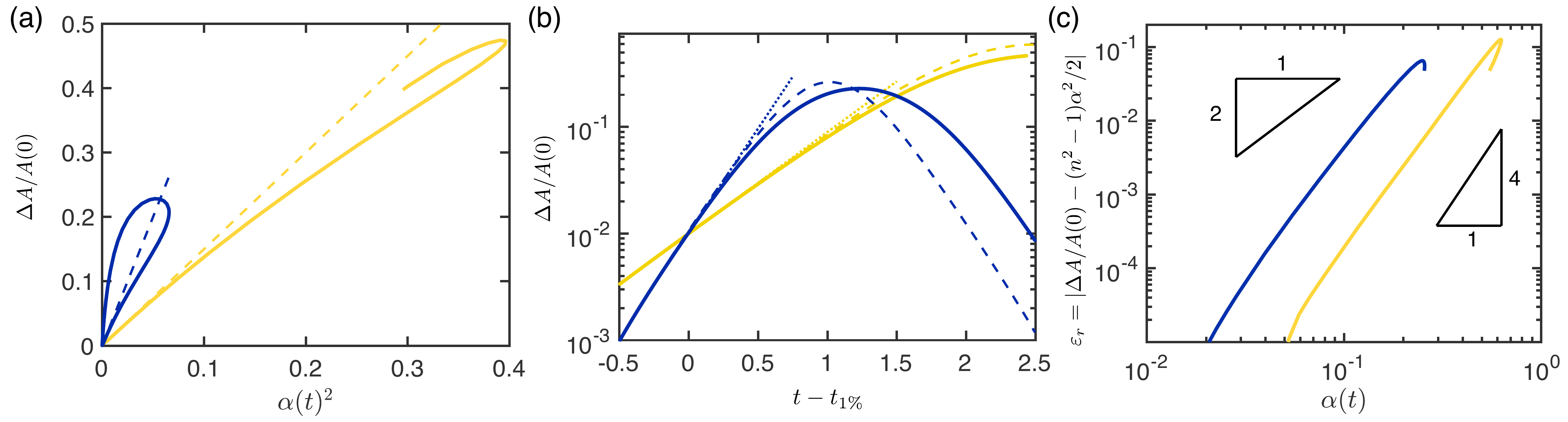}
	\caption{Comparison of the change in area enclosed by the ring  computed numerically and that predicted from the weakly nonlinear analysis presented here. Throughout, results are shown for the $P=3.5$, corresponding to $n=2$, (light/gold curves) and $P=8.7$, corresponding to $n=3$, (dark/blue curves). (a) The relative area change, $\Delta A/A(0)=1-A(t)/A_0$ as a function of $\alpha(t)^2$ shows that for early times the results confirm the  behavior expected from \eqref{eqn:DeltaAPredMSA}, which become the dashed curves in this plot. Numerical results are shown as solid curves, together with the prediction of the weakly nonlinear analysis,  \eqref{eqn:DeltaAPredMSA} (dashed curves). (b) Plotting $\Delta A/A(0)$ as a function of $t$ on semi-logarithmic axes shows that at early times the growth is exponential with growth rate $2\sigma$ where $\sigma$ is as predicted from the linear analysis --- $\Delta A/A(0)\propto\exp(2\sigma t)$ as shown by the dotted lines. (Here times are shown relative to the time, $t_{1\%}$, at which the relative change in area is $1\%$.) (c) A plot of the absolute numerical error, $\mathrm{\varepsilon}_r=|\Delta A/A(0)-(n^2-1)\alpha^2/2|$, as a function of $\alpha(t)$ shows that this error occurs at higher order in $\alpha(t)$ --- most likely $\varepsilon_r(t)\propto\alpha(t)^4$. }
	\label{fig:AreaChange}
\end{figure}

The plots in Fig.~\ref{fig:AreaChange}a show that the prediction of \eqref{eqn:DeltaAPredMSA} is in good agreement with the detailed numerical simulations of the full problem for early times, while the plots in Fig.~\ref{fig:AreaChange}b show that at early times the relative change in area grows according to $\Delta A/A(0)\propto\exp(2\sigma t)$ with $\sigma$ the growth rate of the linear instability. Finally, Fig.~\ref{fig:AreaChange}c shows that the error in the area change occurs at higher order in $\alpha$ than $\mathcal{O}(\alpha^2)$, confirming that the weakly nonlinear analysis presented here is correct to this order.

\paragraph{Compressive force $\lambda$}
The amplitude equation can also be used to study the evolution of  the compressive force $\lambda$ within the ring: the predicted behavior of $\lambda(s,t)$ can readily be computed once the amplitude equation for $\alpha(t)$  \eqref{eq:ampEq} is solved numerically.  

\begin{figure}[h!]
	\centering
		\includegraphics[width=0.75\linewidth]{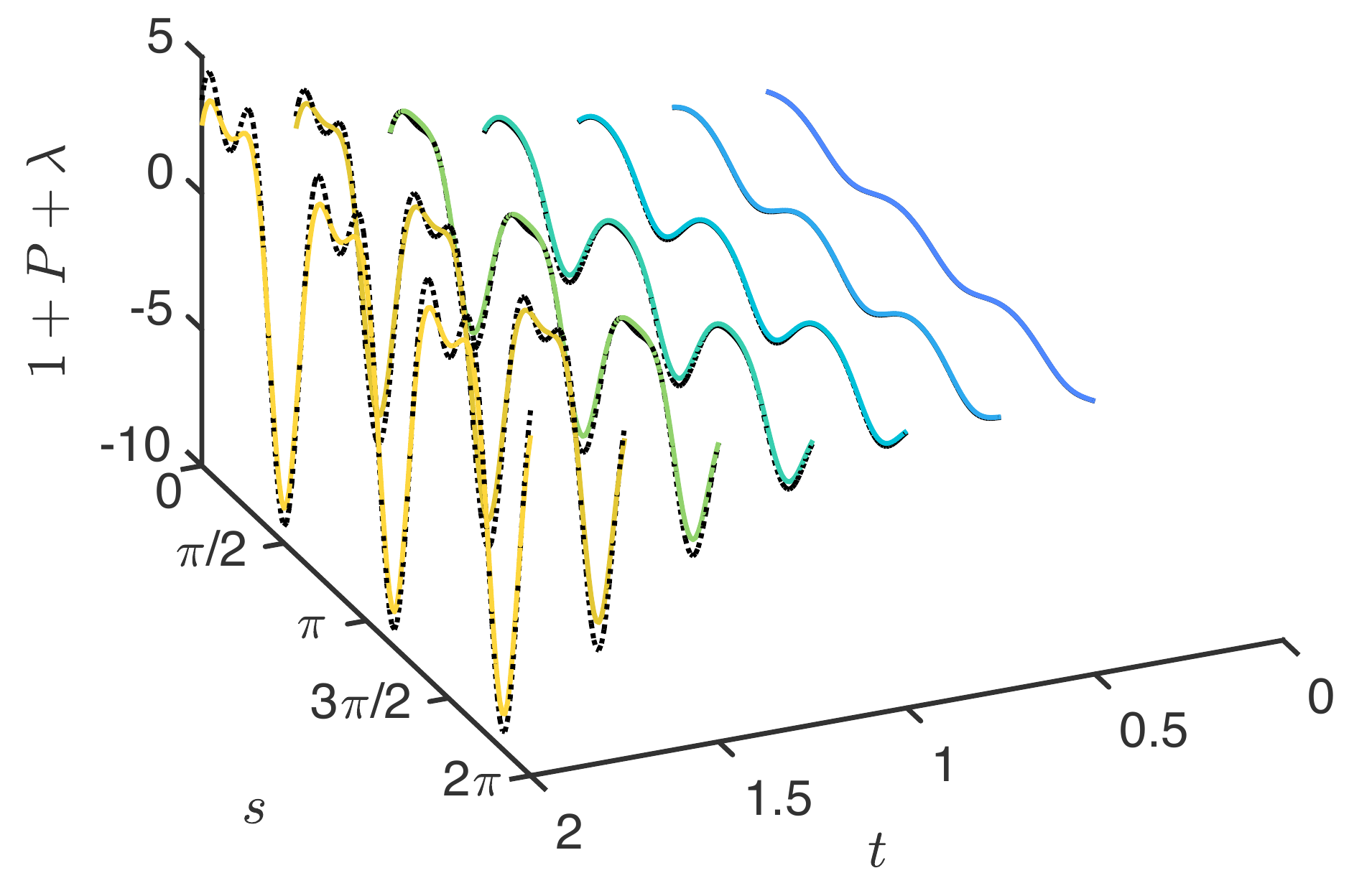}
                \caption{The evolution of the compressive stress within the ring, $\lambda(s,t)$, for an applied pressure $P = 8.7$, corresponding to mode 3 (as seen in Fig.~\ref{fig:MSA-Shape}b). The results of the numerical simulations (solid curves) agree well with the prediction of the multiple scale analysis (dotted black curves). Note, in particular, that at times $t=\mathcal{O}(1)$ the spatial oscillations with arc-length $s$ are not up-down symmetric and that the peaks have a noticeable splitting. These two effects are purely nonlinear effects and are predicted well by the weakly nonlinear analysis presented here. (Results are shown for $1/2\leq t\leq2$ at intervals of $\delta t=1/4$.)}
	\label{fig:LambdaMSA}
\end{figure}

The comparison between the prediction of the weakly nonlinear analysis and full numerical simulations is shown in Fig.~\ref{fig:LambdaMSA}. As with other variables, the comparison between numerics and the weakly nonlinear analysis is very good, particularly at early times. However, this plot of $\lambda(s,t)$ reveals two features that are not so readily observed in other variables: the oscillations in $\lambda(s,t)$ for a fixed time $t$ are not up-down symmetric and, in particular, the crest of these oscillations splits in two, showing the importance of a higher frequency oscillation in the arc-length $s$. Both of these features can be understood by observing from \eqref{eq:U2} that the prefactor of  $\lambda_2\sim n^4\alpha_1^2$ while \eqref{eq:U1} shows that the prefactor of $\lambda_1\sim n^2\alpha_1$. As a result,  $\lambda_2$ is able to compete with $\lambda_1$ despite being strictly at higher order in $\alpha$: the $\cos2ns$ term in the expansion of $\lambda(s,t)$ can  compete with the $\sin ns$ term, breaking up-down symmetry and leading to the splitting of the crests. This is not so apparent in other variables, since the higher order terms remain sub-dominant for longer; for example, $\kappa_2\sim \alpha_1^2n^2$ while $\kappa_1\sim \alpha_1n^2$.

The appearance of second-order variables with frequency $2n$ might suggest this as a precursor of a secondary bifurcation with frequency doubling. Strictly speaking, this is not a secondary instability as these higher modes are present at the primary bifurcation but have very low amplitudes. Further,  we note a competing effect that tends to prevent frequency doubling. Indeed, in the last panel of  Fig.~\ref{fig:NumericalZoology} (with $P=256$), we observe  several events in which two crests merge into a single crest which implies that lobes progressively disappear with time.  We hypothesize that this merger occurs via the `snap-through' of the internal element, as observed in a simple elastic arch \cite{Pandey2014}.
  
\section{Conclusion\label{sec:conclusions}}

We have presented both linear and weakly nonlinear analyses of the dynamic buckling of an elastic ring subject to a suddenly imposed external pressure, $p$. Returning to our initial discussion of the dynamic buckling of a beam under a compressive force ${\cal F}$, we speculated that the same argument might  hold for a ring under pressure $p={\cal F}/a$  and, further, that this is the most important effect of the ring's curvature. This analogy is enough to reproduce the main results of the linear stability analysis for sufficiently large pressures; detailed calculations showed that the most unstable mode is
\beq
n_c\approx \left(\frac{pa^3}{2B}\right)^{1/2},
\label{eqn:nDim}
\eeq which grows with dimensional growth rate
\beq
\sigma_{\mathrm{max}}\approx\frac{1}{2}\frac{pa}{(\rho \Across B)^{1/2}}.
\label{eqn:sigmaDim}
\eeq These large pressure results are precisely the same as those for a beam with ${\cal F}=pa$. We note that while inertia is required to observe buckling at a high mode number, \eqref{eqn:nDim} shows that the mode number selected is independent of the inertia of the ring, $\rho\Across$. Rather the inertia of the ring sets the time scale for the growth of instability, as shown in \eqref{eqn:sigmaDim}. The apparent contradiction that inertia is required to observe a high mode number but is not involved in the selection of a mode number is resolved \cite{BKOCGV} by varying the loading rate  to be below that in \eqref{eqn:sigmaDim}. 

While a linear stability analysis is sufficient to understand the early stages of the dynamics, it does not provide any insight into how the instability develops further. In particular, the linear stability analysis shows that the area enclosed by the loop does not change to leading order; moreover, if one attempts to use its results at higher order, one comes to the obviously erroneous conclusion that the enclosed area \emph{increases} as instability proceeds. Therefore, we performed a weakly nonlinear analysis that allows us to show that the relative change in area $\Delta A/A(0)$ grows exponentially in time, with growth rate $2\sigma_{\mathrm{max}}$. The weakly nonlinear analysis is able to explain other features of our numerical simulations, including an apparent frequency doubling of the compressive force $\lambda$ within the ring. While this might be expected to lead to an increase in the mode number  with time, we note that our numerical simulations show that crests tend to merge with time, presumably due to the increase in confinement that they experience. The time scale of this coarsening remains to be understood.

The dynamic buckling we have studied here may have analogues in other systems of interest. For example, similar dynamics may be relevant in the collapse of veins \cite{chow2006simple} for which it is known that inertial effects due to the fluid--structure interaction are important  \cite{Jensen2003,grotberg2004biofluid}; however, we also note that in these systems axial modes of deformations often dominate the radial modes considered here. Finally, we note that the ring geometry we have studied  may be a more promising paradigm within which to understand dynamic buckling instabilities in tissue and soft materials. While the uniaxial compression of a rod has often been studied \cite{Gladden05}, it can be difficult to impose a known force experimentally and the resulting instability is sensitive to the imposed boundary conditions. The circular symmetry of the ring problem renders the latter issue  irrelevant while our analysis has shown that, up to the instability, the ring curvature translates the normal force given by the applied pressure into a uniform in-plane compression.

\appendix
\section{Details of the numerical scheme}
\label{app:numerics}

In this Appendix we describe the numerical procedure used to determine the evolution of an elastic ring subject to a normal pressure; this motion is governed by the system \eqref{eq:elasticaSystemNonDim}--\eqref{eq:Init_Conditions} and so it is a numerical solution of these equations that we seek. We begin by discussing the discretization of the equations used numerically, before turning to an examination of the constraint associated with the ring's inextensibility.

\subsection{Discretization}

Following \cite{Santillan2007a} we discretize \eqref{eq:elasticaSystemNonDim}--\eqref{eq:Init_Conditions} in  time $t$ \emph{and} in  arc-length $s$; this results in a system of nonlinear equations that are solved using Newton's method. In particular, we consider discrete time points $t_{j} = (j-1) \Delta t$, where typically $ \Delta t = 5 \times10^{-3}$ in the simulations reported in the main paper. The arc-length along the ring is discretized on $N+1$ grid-points at $s_{i} = (i-1) \Delta s $ where $\Delta s = 2\pi/N$, and $i \in [1, N+1]$. 

The system \eqref{eq:elasticaSystemNonDim}--\eqref{eq:Init_Conditions} does not involve any time derivative of the forces $F_{x}$ and $F_{y}$; we  therefore cannot use a simple forward time stepping scheme. Instead, we need to solve for these forces and the rest of the equations together, in a single time step. To do this we use a one-sided time derivative \cite{Chung2002} accurate to $\mathcal{O}(\Delta t^{2})$, namely:

\begin{align}
&\frac{\partial x_i}{\partial t}  = \frac{3 x^{j}_i - 4 x_i^{j-1} + x_i^{j-2}}{2 \Delta t}  = v_{x,i}^j, \label{eq:vxdef}\\
&\frac{\partial y_i}{\partial t}  = \frac{3 y^{j}_i - 4 y_i^{j-1} + y_i^{j-2}}{2 \Delta t}  = v_{y,i}^j. \label{eq:vydef}
\end{align}

Derivatives with respect to arc-length are discretized using central differences. The discretized versions of \eqref{eq:elasticaSystemNonDim}--\eqref{eq:elasticaSystemNonDim_END} read:
\begin{align}
& \frac{x_{i+1}^{j} - x_{i}^j}{\Delta s} = \cos \theta_{i+1/2}^j , \label{eq:a} \\
& \frac{y_{i+1}^j - y_{i}^j}{\Delta s} = \sin \theta_{i+1/2}^j , \label{eq:b}\\
& \frac{\theta_{i+1}^j - \theta_i^j}{\Delta s} = \kappa_{i+1/2}^j , \label{eq:c}\\ 
& \frac{\kappa_{i+1}^j - \kappa_{i}^j}{\Delta s}  = F_{x,\, i+1/2}^j \sin\theta_{i+1/2}^j - F_{y,\,  i+1/2}^j \cos\theta_{i+1/2}^j,    \\
&v_{x,\, i}^{j}=  \frac{3 x_{i} ^{j} - 4 x_{i} ^{j-1} + x_{i} ^{j-2}}{2 \Delta t},\\
& v_{y,\, i}^{j}=  \frac{3 y_{i} ^{j} - 4 y_{i} ^{j-1} + y_{i} ^{j-2}}{2 \Delta t}, \\
&  \frac{F_{x,\, i+1}^{j} - F_{x,\, i}^{j} }{\Delta s}  - P \sin \theta_{i+1/2} =\frac{3 v_{x,\, i+1/2} ^{j} - 4 v_{x,\, i+1/2} ^{j-1} + v_{x,\, i+1/2} ^{j-2}}{2 \Delta t}  ,   \\
&  \frac{F_{y,\, i+1}^{j} - F_{y,\, i}^{j}}{\Delta s}  + P \cos \theta_{i+1/2}^j = \frac{3 v_{y,\, i+1/2} ^{j} - 4 v_{y,\, i+1/2} ^{j-1} + v_{y,\, i+1/2} ^{j-2}}{2 \Delta t} . \label{eq:z}  
\end{align} Note that here we use the notational convention that $f_{i+1/2}=(f_i+f_{i+1})/2$ for $f\in\{v_x,v_y,\theta,\kappa,F_x,F_y\}$.

Given the values of the variables at times $t_{j-2}$ and $t_{j-1}$, the system \eqref{eq:a} -- \eqref{eq:z} can  be solved at time $t_j$ using a Newton solver; we use the MATLAB routine \texttt{fsolve} to find these solutions.\\

\subsection{Inextensibility constraint}

\label{app:accuracy}

Here we discuss the constraint of inextensibility focussing, in particular, on whether the total length of the ring is preserved over time during the integration. Our approach is intended to ensure that the inextensibility constraints \eqref{eq:elasticaSystemNonDim}--\eqref{eq:constraintY} are automatically satisfied; this is achieved by solving \eqref{eq:a}--\eqref{eq:b} together with the remaining equations \eqref{eq:c}-\eqref{eq:z} at each time step. 

In the following, we check that the inextensibility constraint is satisfied for the simulations reported in Fig.~\ref{fig:NumericalZoology}.
To do this, we numerically determine the arc-length of the ring as 
\begin{align}
	\ell(t_j) =  \sum_{i=0}^N\bigl[(x^j_{i+1}-x^j_{i})^2+(y^j_{i+1}-y^j_{i})^2\bigr]^{1/2}.
	\label{eq:elltj}
\end{align}

\begin{figure}
	\centering
	\includegraphics[width=\linewidth]{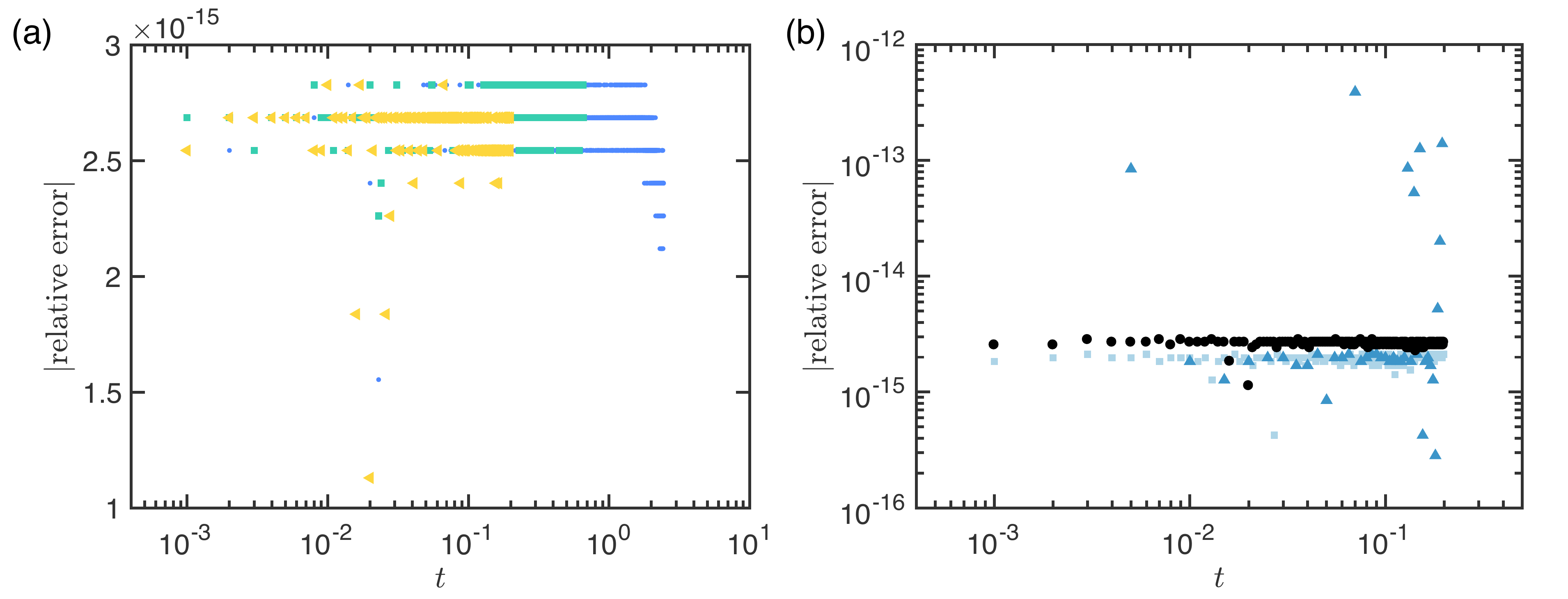}
	\caption{Evolution of the relative error in the total arc-length of the ring, $\ell(t_j)$, defined in \eqref{eq:elltj},  as computed for the numerical simulations presented in Fig.~\ref{fig:NumericalZoology}. Here, we plot the relative error, $|(\ell(t_j)-2\pi)/2\pi|$, as a function of time for different driving pressures and resolutions.
			(a) With fixed temporal and spatial resolution, $\Delta t = 10^{-3}$ and $N = 256$, results are shown for different pressures $P = 8$ (dots), $P = 32$ (squares) and $P = 128$ (left-pointing triangles). 	
			(b) With fixed pressure, $P=128$,  simulation results are shown with different spatial and temporal resolutions as follows: $N=128$ and $\Delta t = 5\times10^{-3}$ (triangles), $N=128$ and $\Delta t = 10^{-3}$ (squares), $N = 256$ and $\Delta t = 10^{-3}$(dots). In all these simulations, the magnitude of the relative error remains below $10^{-12}$ throughout.} 
	\label{fig:Constraint_ElasticRing}
\end{figure}

In Fig.~\ref{fig:Constraint_ElasticRing}, we plot the magnitude of the relative error in the total arc-length of the ring, $|(\ell(t_j) - 2\pi)/2\pi|$, as a function of time for different values of the driving pressure and for different spatio-temporal resolutions.
Fig.~\ref{fig:Constraint_ElasticRing}a shows simulations for different values of pressures, $P=8, 32, 128$, at a fixed value of the time step and number of grid points ($\Delta t=10^{-3}$ and $N=256$). In contrast, Fig.~\ref{fig:Constraint_ElasticRing}b shows the relative error for simulations with a fixed pressure, $P = 128$, but different spatial and temporal resolutions. We see here that the ring does indeed maintain its length over time as guaranteed by our numerical scheme --- the magnitude of the relative error in the total length for the simulations reported in Fig.~\ref{fig:NumericalZoology} remains less than $10^{-12}$ throughout.

Finally, we discuss two limitations of our numerical scheme. First, the initial conditions we used, \textit{i.e.} \eqref{eq:Init_Conditions}, contains a random noise at each spatial grid point. This means that the initial condition is mesh dependent. However during the simulations reported in Fig.~\ref{fig:NumericalZoology} we noticed that, at the first time step, the solutions initially converge to a circular ring shape, and then the fastest growing modes start emerging. This suggests that the observed instability is an intrinsic feature of the solutions of \eqref{eq:elasticaSystemNonDim} -\eqref{eq:elasticaSystemNonDim_END}. Second, it is to be noted that since we have used a \textsc{BDF2} (Backward Differentiation Formula) in \eqref{eq:vxdef} and \eqref{eq:vydef}, it follows that our scheme will be dissipative, and only accurate at $\mathcal{O}(\Delta t^2)$. This limitation implies that the solutions reported in this paper cannot be expected to be accurate at very late time --- a limitation that might be overcome by using a symplectic integrator or, in general, a geometric integrator \cite{Marsden1998}.

\section{Details of the weakly nonlinear analysis}
\label{app:wna}

In this Appendix, we provide more details of the weakly nonlinear analysis that is performed to describe the behavior when the pressure is just above the critical pressure of buckling of the $n^{\text{th}}$ mode.
We introduce $ \bm{U} =[u,v,\phi,\kappa,\lambda]^{T}$ and  linearize  equations \eqref{eq:GOE-Polar-Start}--\eqref{eq:GOE-Polar-End} around the static solution $ \bm{U}_{0} =[0,0,0,1,-P-1]^{T}$ by computing $\bm{U}=\bm{U}_{0}+\epsilon \bm{U}_{1}$ to first order. We obtain a linear homogeneous system of the form 
\begin{equation}
\mathcal{L}_{P} \bm{U}_1 = 0,
\end{equation}
 where $\mathcal{L}_{P}$ is a linear differential operator in $s$ and $t$ with constant coefficients and is dependent on the control parameter $P$. The solution of this equation is given by~(\ref{eq:U1}) with $\alpha_{1}(t)=\exp(\sigma t)$. Here, the growth rate $\sigma$ is given by  the dispersion relation depending on the  mode $n$, and  pressure $P$: $\sigma = \sigma(n,P)$. We define the critical pressure $p_{n}$ by the first positive root of  $\sigma(n,p_{n})=0$, which leads to $p_n = n^2-1$. The leading-order problem therefore precisely reproduces the results of the linear stability analysis, as it must.

Next, we perform the weakly nonlinear analysis. We assume that the pressure  $P = p_n+\epsilon^2$ is slightly above the critical pressure $p_n$.  Doing so, we relate the distance to the bifurcation to a small parameter $\epsilon$. Furthermore, since we place ourselves at the bifurcation, the  only dependence on time is through a slow time scale. An analysis of the dispersion relation shows that the relevant growth rate is $O(\epsilon)$ and hence we expect evolution to occur on a time scale $\tau=\epsilon t$. This motivates the expansion 
\begin{align}
\bm{U} = \bm{U}_0 + \epsilon \bm{U}_1 + \epsilon^2 \bm{U}_2 + \epsilon^3 \bm{U}_3 + \mathcal{O}(\epsilon^4), \label{eqn:Uexpansion}
\end{align}
where $\bm{U}_0 $ is as before and $
      \bm{U}_j =[
           u_j(s,\epsilon t) ,
           v_j(s,\epsilon t) ,
           \phi_j(s,\epsilon t) ,
           \kappa_j(s,\epsilon t),
           \lambda_j(s,\epsilon t)]^{T}$.
Substituting  this expansion  into the governing equations \eqref{eq:GOE-Polar-Start}--\eqref{eq:GOE-Polar-End} results in a hierarchy of systems for the $ \bm{U}_j$. To first order, the system reads $\mathcal{L}_{p_{n}} \bm{U}_1= 0$ and due to the particular choice of $P$, we obtain~(\ref{eq:U1}) where  $P=p_{n}$  and $\alpha_{1}(\tau)$ is now arbitrary.

To second order, we obtain a system of the form
\begin{align}
\mathcal{L}_{p_{n}}  \bm{U}_2 = \mathbf{G}_{2}(\bm{U}_1)
\end{align}
whose unknown is $ \bm{U}_2$ and where $\mathbf{G}_{2}(\bm{U}_1)$ is a  function of $\bm{U}_1$. The solution of this system is
\begin{align}
\bm{U}_{2} &= 
\frac{\alpha_{1}^2(\tau)}{8n^3}\begin{bmatrix}
 2n(n^2-1)^2\\
-(n^2-1)^2\sin(2 n s)\\
-2n(n^2-1)^2\cos(2 n s) \\
 - (n^2-1)^2\sin(2 n s)\\
2n\frac{(n^4-1)^2}{(n+1)^2}\cos(2 n s)
    \end{bmatrix}.
\end{align} 

To third order, the system for $ \bm{U}_3$ reads
\begin{align}
\mathcal{L}_{p_{n}}  \bm{U}_3= \mathbf{G}_{3}(\bm{U}_1,\bm{U}_2,\alpha_{1,\tau\tau}),
\end{align}
where the extra dependence on the second derivative of $\alpha_{1}(\tau)$, $\alpha_{1,\tau\tau}$, comes from the assumption of the long time scale. In general, this system does not have a solution as the inhomogenous term is not in the image of the operator $\mathcal{L}_{p_{n}}$. Therefore, a compatibility  condition, the so-called Fredholm alternative \cite{gonita99, Neukirch2012},  needs to be used. Let $\boldsymbol{\xi}$ be the solution of the homogeneous adjoint problem $\mathcal{L}_{p_n}^\ast \boldsymbol{\xi}=\mathbf{0}$. Then, this compatibility condition to third order simply reads
\begin{equation}
 \boldsymbol{\xi} \cdot \mathbf{G}_{3}(\bm{U}_1,\bm{U}_2,\alpha_{1,\tau\tau})=0.
\end{equation}
This condition provides a differential equation for $\alpha_{1}$. Taking into account the different change of variable and defining $\alpha(t)=\epsilon \alpha_{1}(\epsilon t)$, we obtain the amplitude equation~(\ref{eq:ampEq}).

Once $\bm{U}$ is known to second order, we may use the relations \eqref{eq:R} to express the position of the ring in Cartesian coordinates. Accurate to $\mathcal{O}(\epsilon^2)$ we find that:
\begin{align}
&x(s,t) = \sin(s)-\alpha(t) \left(\sin s \sin n s+\frac{\cos s \cos n s}{n}\right)-
   \alpha^2(t) \left[\frac{\left(n^2-1\right)^2 }{4n^2}\sin s+\frac{\left(n^2-1\right)^2}{8 n^3}\cos s \sin 2ns\right],\\
&y(s,t) = -\cos s+ \alpha(t) \left(\cos s \sin ns-\frac{\sin s \cos ns}{n}\right) + \alpha^2(t) \left[\frac{\left(n^2-1\right)^2 }{4
   n^2}\cos s-\frac{\left(n^2-1\right)^2}{8 n^3} \sin s\sin2ns\right].
   \label{eq:yst}
\end{align} 

We may use \eqref{eq:Curvature} to express $\theta(s,t)$ which is given by:
\begin{align}
\theta(s,t) = s+  \frac{n^2-1}{n} \alpha(t)\left(   \cos ns-1\right)-\frac{ \left(n^2-1\right)^2}{8 n^3}\alpha^2(t)\sin 2n s.
\end{align}

Similarly, the expression for the internal forces may be expressed in Cartesian coordinates using \eqref{eq:Force} which is given component-wise as:
\begin{align}
&F_x = \frac{\partial^2 \theta}{\partial s^2} \sin\theta+ \left(\frac{\partial \theta}{\partial s}\right)^2 \cos\theta+\lambda\cos\theta, \\
&F_y = -\frac{\partial^2 \theta}{\partial s^2}  \cos\theta+\left(\frac{\partial \theta}{\partial s}\right)^2 \sin\theta+\lambda\sin\theta.
\end{align}

\begin{acknowledgments}
 \noindent\textbf{Acknowledgments--\ }
The  research leading to these results has received funding from the European Research Council under the European Union's Horizon 2020 Programme/ERC Grant Agreement no.~637334 (DV) and the Engineering and Physical Sciences Research Council grant EP/R020205/1 (AG). We are grateful to Finn Box for comments on an earlier version of this article and to Peter Howell for discussions about dynamic buckling that led to this work.  
\end{acknowledgments}

\end{document}